# 3D-Printed Terahertz Topological Waveguides

Muhammad Talal Ali Khan[1†*], Haisu Li[2 †*], Nathan Nam Minh Duong[1], Andrea Blanco-Redondo [3], And Shaghik Atakaramians[1*]

[1]School of Electrical Engineering and Telecommunications, UNSW Sydney, New South Wales 2052, Australia

[2]Key Laboratory of All Optical Network and Advanced Telecommunication Network of EMC, Institute of Lightwave Technology, Beijing Jiaotong University, Beijing 100044, China

[3]Nokia Bell Labs, 791 Holmdel Rd, Holmdel, NJ 07733, USA

[†]M.T.A.K. and H. L. contributed equally to this work.

[*]Correspondence to: m.t.khan@unsw.edu.au , lihaisu@bjtu.edu.cn , s.atakaramians@unsw.edu.au.

(Date: September 21st , 2020)

**Compact and robust waveguide chips are crucial for new integrated terahertz applications, such as high-speed interconnections between processors and broadband short-range wireless communications. Progress on topological photonic crystals shows potential to improve integrated terahertz systems that suffer from high losses around sharp bends. Robust terahertz topological transport through sharp bends on a silicon chip has been recently reported over a relatively narrow bandwidth. Here, we report the experimental demonstration of topological terahertz planar air-channel metallic waveguides which can be integrated into an on-chip interconnect. Our platform can be fabricated by a simple, cost-effective technique combining 3D-printing and gold-sputtering. The relative size of the measured topological bandgap is ∼12.5%, which entails significant improvement over all-silicon terahertz topological waveguides (∼7.8%). We further demonstrate robust THz propagation around defects and delay lines. Our work provides a promising path towards compact integrated terahertz devices as a next frontier for terahertz wireless communications.**

Terahertz (THz) radiation has proven crucial in a wide range of applications including broadband wireless communication, biology and chemical sciences, non-destructive evaluation and spectroscopic sensing[1]. For instance, THz wireless communication has the potential to enable seamless interconnection between ultra-high speed wired networks and personal wireless devices, as well as achieving terabit-per-second data rates for high bandwidth applications[2,3]. This breadth of applications is driving research for integrated and cost-effective on-chip THz technology. In order to build low-loss, low-dispersion and low-cost THz integrated circuits, the development of compact and robust THz waveguides is of particular interest, mainly in the lower part of the spectrum (∼ 0.1 − 1 THz)[4,5]. Planar waveguides, such as transmission lines[6,7], parallel-plate[8,9], dielectric[10,11] and photonic crystal[12-14] waveguides have been proposed as promising approaches for THz applications, thanks to their potential of integration with more functional devices in contrast to the fibre-type THz waveguides[15-18]. Nonetheless, these platforms do not support low-loss sharp bends[19,20] and suffer from loss related to fabrication imperfections,[13] which complicates their use in practical device designs.

In parallel, the discovery of topological insulators in condensed matter[21-23] has opened the gate to fascinating features like unidirectional propagation of electronic edge states that are robust against imperfections and disorders. In 2008, these ideas were translated to the field of photonics[24,25], which spurred a wave of research on new platforms that supported topologically protected light propagation[26-30]. This research has opened new avenues for topologically protected on-chip optical devices, such as delay lines[26], topological lasers[31,32], slow-light waveguides[33] and quantum circuits[34-38]. Most of the aforementioned developments have been demonstrated at microwave[25,39-41] and optical frequencies[30,31,34,36,42-45]. Recently, Yang et al.[46] demonstrated robust transport of THz signals in an all-silicon waveguide resembling a valley Hall photonic topological insulator (PTI). Such waveguide can offer low propagation losses of 0.05 dB/mm within a relatively narrow topological bandgap of 0.322 − 0.348 THz (relative bandwidth of ~7.8%). Moreover, the mass production of complex THz device remains challenging when it comes to the conventional intensive fabrication methods like micromachining/photolithography[47,48]. Fortunately, recent developments in 3D printing technologies[49] offered the possibility to increase the compatibility of on-chip integration[50] and paved the way to fabricating complex structures at low cost and smaller feature sizes[47,51,52] in the frequency range from 0.1 THz to ~1 THz [50,53-55].

Here, we leverage the idea of the bianisotropic metawaveguide, theoretically proposed by Ma et al.[20], to produce an air-channel THz topological integrated waveguide with an enhanced bandwidth and low fabrication cost. Our platform is fully fabricated by 3D printing and gold-sputtering. We experimentally investigate topological waveguides based on quantum spin Hall (QSH) effect, including straight paths with different lengths, a defect path and a delay line in the THz range. Our measurements demonstrate that the topologically protected propagating mode only exists when bianisotropy is introduced in the waveguide while no mode is guided when the bianisotropy is eliminated. Comprehensive experimental characterization confirms a single-mode, linear-dispersion transmission window within the topological photonic bandgap between 0.143 THz to 0.162 THz (relative bandwidth of 12.5% compared to 7.8% in Ref. 46). Moreover, we demonstrate the robustness of topologically protected propagation along the defect path and the delay line according to the measured phase information.

**Results**
**Design and Fabrication**. Figure 1 (a) presents a schematic of the THz photonic topological waveguide where two-dimensional hexagonally arranged photonic crystals composed of metallic pillars are attached to one of the two confining parallel metallic plates. A symmetry is broken by introducing the finite gap $g$ between the parallel plates and the pillars. Chiral-dependent unidirectional propagation of THz signals is supported through an air channel at the interface of two sections of the hexagonal lattice – one with an airgap between the pillars and the top plate, and the other with the gap between the pillars and the bottom plate. Due to a bianisotropic response, the structure behaves as a QSH-PTI[20,39] with a topological bandgap, and it is expected to support unidirectional propagation of opposite circular polarizations along the +z and -z directions, respectively.

To achieve a complete topological photonic bandgap at the target frequency of 0.15 THz, we design a hexagonal unit cell with lattice constant $a$ = 1500 µm, pillars' diameter at the middle height $d$=500 µm, pillars' height $h$=1250 µm, and gap $g$=200 µm [inset of Fig. 1 (a), front view]. The numerical simulations of the unit cell confirm the existence of the topological bandgap (see Methods for more details on numerical models). Since the pillar walls will have to be angled to ensure even sputtering during fabrication, we calculate the band structure, with 10° and 20° tilt angles on the pillar walls, and with straight pillars [Fig. 1 (b)], where $d$ is invariant. The results show that the topological bandgap stays open around the $K$ corner of the Brillouin zone, and even increasing the tilt angle of pillars enhances the topological bandgap width to 9 GHz and 9.5 GHz for 10° and 20° tilts, respectively [see the zoom-in inset in the bottom-right of Fig. 1 (b)]. Next, we carry out a supercell simulation to optimize the gap size (note that other parameters such as $h$ are invariant). From Fig. 1 (c), we observe no transmission for $g$ = 0 (which will be experimentally verified later). With an increase of $g$, the bandwidth improves. The maximum bandgap width for mode guidance could be up to 19 GHz (relative bandwidth of 12.5%) when the gap size $g$ is between 122 – 207 µm. The only effect we can observe is the shift of the guidance window if the gap size deviates from its designed dimension i.e. 200 µm. This high tolerance to variations in the gap size ensures that the structure is robust to fabrication imperfections. For $g$ > 207 µm, the topologically protected transmission range narrows due to weakened bianisotropy responses.

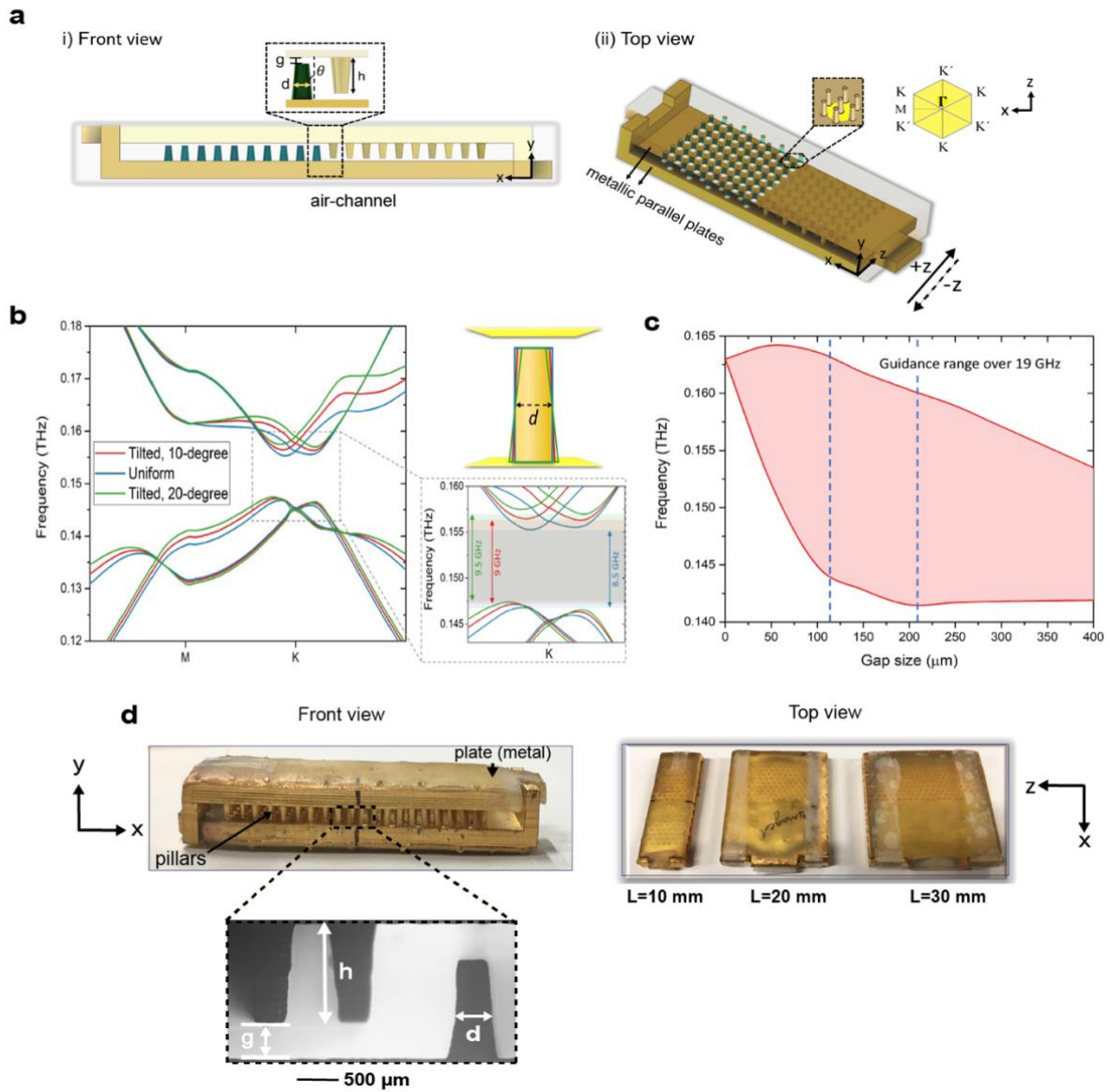

**Figure 1. Design and fabrication of the THz metallic topological photonic crystal waveguide. a,** Schematic of the THz photonic crystal waveguide. (i) Front view featuring the enclosure of the pillars between parallel metallic plates. The enlarged region on the top defines the geometric parameters of the waveguide. (ii) Top view: The inset depicts the Brillouin zone and the triangular lattice. The unidirectional mode is guided along the MK edge of the Brillouin zone. **b,** Projected photonic bandgap diagram with and without tilt angles. Top-right inset shows the schematic with the tilt angles (10°, red) and (20°, green) that are introduced to achieve even coating during fabrication. Bottom-right inset shows the band diagrams around *K* point. **c,** Mapping of the frequency range of guided modes as a function of gap size in the range of 0 to 400 μm. The region in between the dashed lines illustrates the maximum guidance range of around 19 GHz. **d,** Photographs of the assembled fabricated samples. The bottom inset shows the microscope image of the front view, highlights the parameters and the air-channel with two opposing pillars. The top view displays three different lengths (*L*=10, 20, 30 mm) of the straight path waveguides.

Based on the designed parameters described above, we fabricated several samples of topological waveguides, including straight paths with different lengths, a straight path with a defect (two pillars missing, one attached to each metal plate), and a delay line (consisting of chained 60-degree and 120-degree bends). The samples were fabricated using a 3D printing technique[49] followed by gold sputtering (see Methods for detailed fabrication process). Such a simple, flexible and cost-effective fabrication approach suggests that mass production of the THz topological waveguides could be possible. The straight path fabricated samples are shown in Fig. 1 (d). The measured average parameters of the fabricated waveguides are as follows: $d$ = 500 $\pm$ 30 µm, $h$ = 1250 $\pm$ 30 µm and $g$ = 200 $\pm$ 10 µm, which, overall, satisfy our design values. To check the influence of imperfect pillars parameters within acceptable tolerance, we perform supercell simulations on different pillar dimensions including diameter and height. The results confirm the increase in measured losses and reduction in topological bandgap width (we discuss fabrication tolerance in the section of Dispersion and Loss Characterization; for details of the simulation results, see Supplementary Note 1, Section B). In the next section, we will experimentally verify the topological protection phenomenon.

**Verification of Topologically Protected Propagation.** We employ a commercial THz-Time domain spectroscopy (THz-TDS) system to experimentally demonstrate the topological protection in our platform. Note that the THz source and detector are linearly polarized, oriented along the *y* direction (i.e., perpendicular to the plates, see Methods and Fig. S3). We first verify the existence of two counterpropagating topologically protected guided modes in the waveguide. To do so, we measure the transmission in the presence of the gaps that induce bianisotropy. Subsequently, we close the gaps between the pillars and plates and measure the transmission in the absence of gaps. Figure 2 (a) shows the normalized electric-field amplitudes for both forward (+z) and backward (-z) propagating directions. We observe transmission through the straight path waveguide when the gap is open (solid curves) and no transmission when the gap is closed (dashed curves). Moreover, the loss observed on the measured spectra of the topologically guided modes can be related to mode mismatch between a linearly polarized incoming THz beam and the circularly polarized topological waveguide mode. The numerical simulation indicates a coupling loss around 10 dB in total; into and out of the waveguide (detailed discussion of waveguide coupling in Supplementary Note 2).

Next, to confirm the transmission indeed occurs due to the guided mode localized in the air-channel, we measure the electric field amplitude in the far-field. The raster scan measurement was performed on a 2.00 mm x 1.45 mm area (step size of 0.5 mm in *xy*-plane) with an aperture of 1.50 mm at the output facet of the waveguide for imaging purposes (see Methods for more details). The far-field maps of the electric-field amplitude is shown in Fig. 2 (b) and (c) with (*g* = 200 μm) and without (*g* = 0 μm) gap, respectively. The amplitude maps at 0.152 THz (center frequency) further illustrate that the excited mode is indeed guided within the air-channel when the pillar-plate air-gaps exist for both $E_x$ and $E_y$ components of the mode [see Fig. 2 (b)]. Notably, we rotated the detector head to measure the $E_y$ ($\pi/2$ phase difference with respect to $E_x$ ) component. The corresponding well-matched numerical simulation results are shown in Fig. 2 (d). In short, the experiments confirm the existence of the bandgap and mode propagation in the presence of the air-gaps, while no transmitted mode is observed for the gapless configuration. The measurements demonstrate topologically protected propagation over the 0.143 – 0.162 THz range.

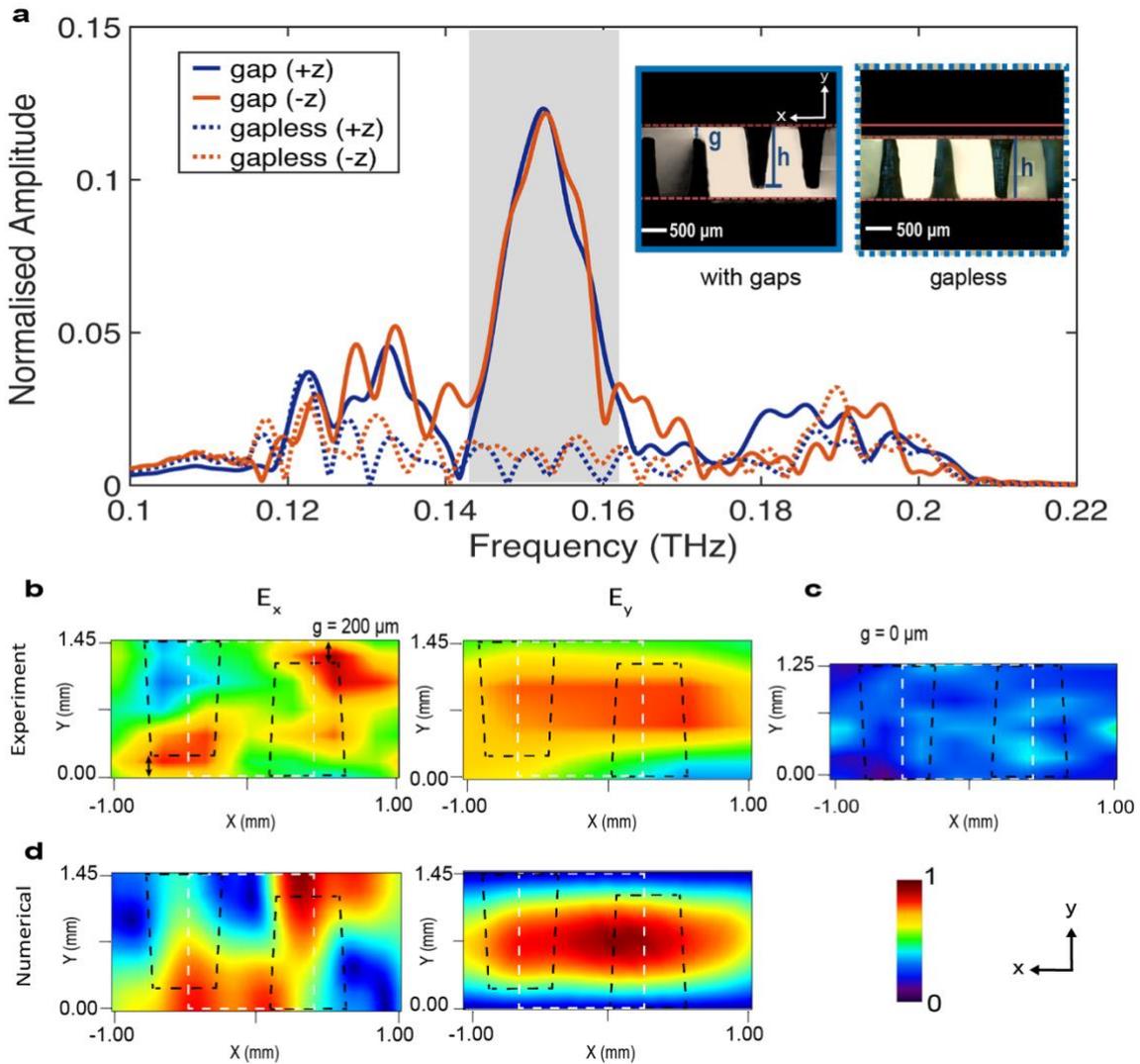

**Figure 2. Experimental validation of topologically protected modes in the fabricated metallic THz waveguide. a,** Normalized field amplitude measured from a straight-path waveguide with and without gaps. The grey area represents the measured bandgap region (0.143 – 0.162 THz). Two solid transmission peaks correspond to mode guidance in the presence of gaps (*g* =200 μm) for forward (+*z*) and backward (-*z*) propagation directions while dashed curves show the absence of guided mode when the gaps are fully closed (*g* = 0 μm). The insets show the measured electron microscopy images of the assembled sample with gaps (left) and gapless (right) configurations. **b–c,** Experimentally normalised amplitudes of electric field (at 0.152 THz) measured for -*z* propagation, when **b** g= 200 $\mu$m for $E_x$ and $E_y$ components of the mode and **c** g = 0 $\mu$m. The raster scan is conducted in an area (2.00mm × 1.45mm), where the black dashed lines mimic the pillars adjacent to the air-channel (white dashed block). The field profile illustrates the localization of the guided mode in the topological channel. **d,** Corresponding field maps from simulations.

**Dispersion and Loss Characterization.** We characterize the waveguide dispersion and loss using a set of identically processed waveguides with three different lengths to emulate the cut-back method. We measure the electric field amplitudes at the output of 10 mm, 25 mm and 30 mm straight-path waveguides, as shown in Fig. 3 (a). The schematic of the straight path, including the direction of propagation of both circular polarizations, is presented in the inset of Fig. 3 (a). As expected, high transmission occurred within the frequencies of bandgap region (0.143 – 0.162 THz) for all three path-lengths. The 3-dB bandwidth is approximately 8 GHz (10 mm), 7.5 GHz (25 mm) and 7 GHz (30 mm). Fig. 3 (b) shows the normalized dispersion, where the solid (+z) and dashed (-z) curves represent the two circularly polarized modes, respectively. The error bars show the standard deviation of the measurements. The experimental dispersion curves show two unidirectional modes that cross at the center of the bandgap. Moreover, the guided modes should have a linear slope in the bandgap [Fig. 1 (b)] as expected from simulations and the nature of these topological edge modes. In order to check the linearity of the measured dispersion curves, we fit linear dispersion curves and find out that the $R^2$ values are 99.9% and 96.3% for +z and -z propagation, respectively. This implies near zero group velocity dispersion (i.e., the second derivative). Then, we compare the experimental (averaged) and theoretical propagation losses in Fig. 3 (c). The measured loss within the bandgap is < 0.3 dB/mm for both +z (solid red line) and -z (blue dashed line) propagation directions, contrasting with loss < 0.2 dB/mm in the simulations (solid and dashed black lines). We further notice a narrower transmission band in the experimental characterization. To explain this, we numerically investigate the effect of both pillar's diameter and height of the fabricated waveguide on the loss (see Fig. S2, Supplementary Note 1). We observe that an undershoot of the pillar diameter (-50 µm) increases the loss for forward propagation at low frequencies (0.145 – 0.150 THz) and backward propagation at high frequencies (0.152 – 0.162 THz). This is while the overshoot of the pillar diameter (+50 μm) has negligible effect on the loss. The influence of pillar height on propagation loss is minor relative to the pillar diameter – a slight increase of losses at the low frequency limit of the bandgap for an overshoot of pillar height (+50 μm). Another reason for higher measured loss values could be the slight misalignments of the waveguide with the incoming beam. A 100 μm misalignment due to the shift in beam spot center along the *x*- and *y*-axis introduces additional coupling loss around 0.1 dB for each input and output ends (for more details, see Supplementary Note 2), which matches the discrepancy observed. This can easily happen as we are measuring three different lengths and are aligning the waveguides to get maximum transmission. Furthermore, our supercell simulations indicate that waveguides with more pillars in the cladding can reduce the loss further, in particular at frequencies close to the bandgap edge (see Fig. S1, Supplementary Note 1).

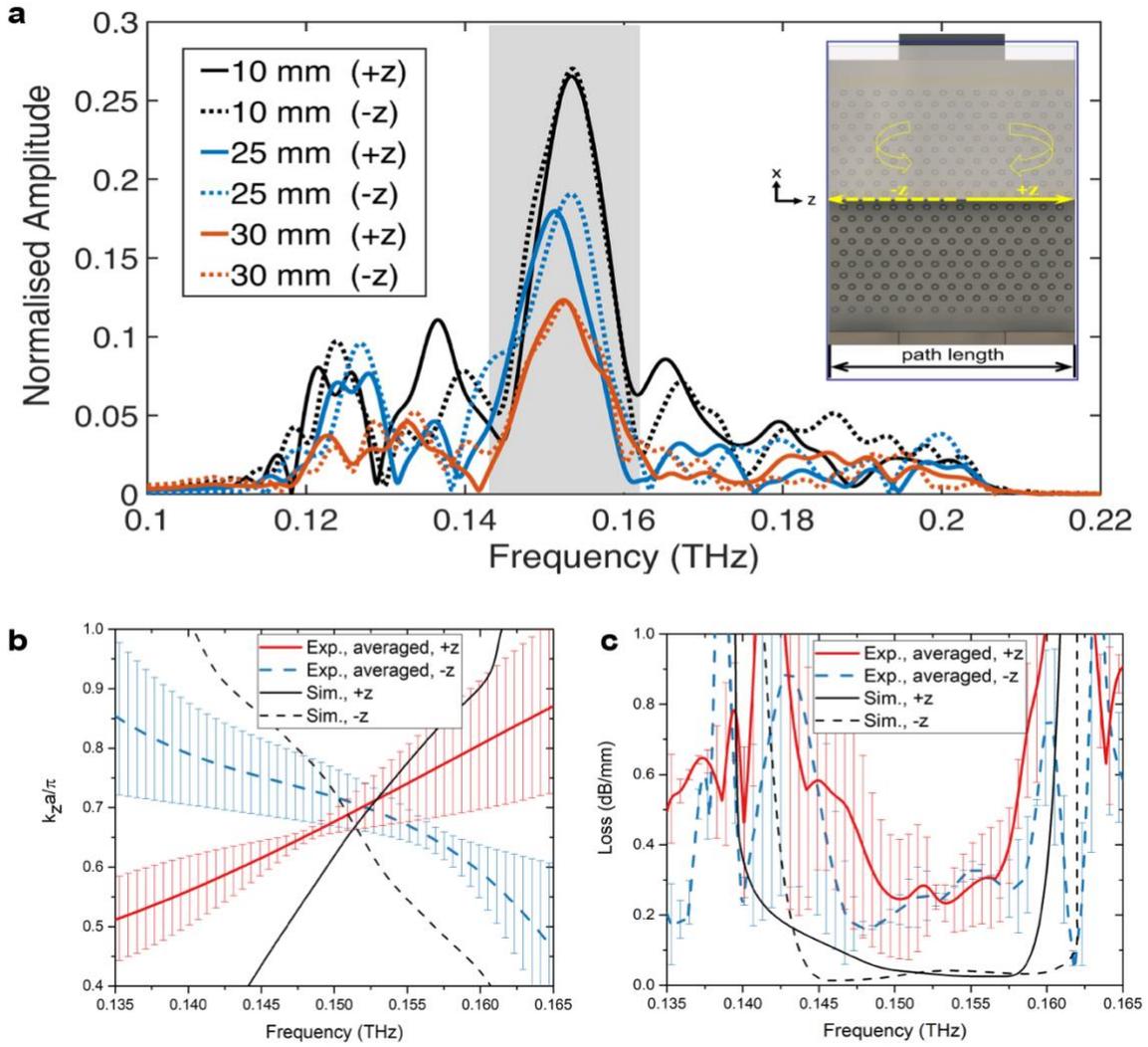

**Figure 3. Experimental characterizations of three different straight-path length waveguides and properties of the waveguided mode. a**, Measured electric filed amplitude curves for 10 mm, 25 mm and 30 mm straight-path waveguides using THz-TDS experimental setup. The shaded frequency range shows the measured bandgap from 0.143 to 0.162 THz. The transmissions are denoted by the solid (+z) and dashed (-z) lines. The inset displays the straight path schematic and the propagation of circularly polarized modes along the topologically protected channel with opposite signs. Note that half-top metal plate is removed to expose the pillars attached to the bottom plate. **b,** Dispersion relations of the channel. The curves present both experimental (averaged) and numerical super-cell simulation results of spin-polarized surface waves supported by the topologically protected channel. **c**, Propagation losses of the waveguide for both experimental and theoretical results in the overall bandgap region. The error bars indicate the standard deviations of the results.

**THz Topological Robust Transport**. Here, we present THz topological robust transport in the forward (+*z*) direction through a defect path and through a delay line consisting of a path with sharp bends [see Supplementary Note 3 for backward (-*z*) propagation results]. First, we concentrate on a structure in which two pillars (one from each parallel plate) are removed from the centre of the waveguide path. The schematic of one side of the waveguide is shown in the top panel of Fig. 4 (a). Figure 4 (b) presents the measured normalized electric field amplitude of the straight path (20 mm path length) with and without defects. The measurements demonstrate that the structure with defects shows almost identical transmission profile to the perfect straight waveguide. Following, we characterize the propagation of light along an interface with sharp bends. The addition of sharp bends increases the propagation length, providing a way to introduce delay in the transmitted light[39]. The interface contains four sharp corners including two 120° and two 60° turns [see Fig. 4 (a) bottom panel]. To quantitively prove the topological protection, we compare the amplitudes measured from delay line (total path length of 32 mm including sharp bends) with the straight path (30 mm path length without bends) as presented in Fig. 4 (c). The measured transmission is very similar in both cases, highlighting once again the topological robustness exhibited by our platform. The inset of Fig. 4 (c) depicts the simulated energy density distribution of the delay line at 0.15 THz illustrating uniform field amplitude, as it propagates along the channel regardless of bends. Note that the wave is launched from the left side (red arrow) in the delay line path. Hence, we can expect that the THz waves travel along the bends in absence of Fabry-Perot resonances that generally happens for conventional photonic crystal waveguides[20,56].

To further verify the robustness of the THz transport, we examine the phase profiles and time delays difference through paths with the defect and the delay line. The measured phases are plotted in Fig. 4 (d) and (e) as a function of frequency. In addition, we fit a linear curve for the measured phase data within the transmission window. As shown in Fig. 4 (d) the relative phase trends of the 20-mm long straight path with and without defects are virtually parallel and overlap. This illustrates that the presence of the missing pillars does not affect the topologically protected propagation. Figure 4 (e) shows that the relative phases of the delay line and the straight path (blue curve, 30 mm) overlap. Note that the relative phase of 30 mm straight path is shown for a narrower frequency ranges (0.148 – 0.152 THz) due to higher noise levels. Then, we calculate the time delay difference between two waveguides as follows[39],

$$\tau = \frac{1}{2\pi} \left( \frac{\partial \varphi_{(defect/delay/30\text{mm})}}{\partial f} - \frac{\partial \varphi_{(20\text{ mm})}}{\partial f} \right) \quad (1)$$

where $\varphi_{(defect/delay/30\text{mm})}$ is the relative phase of the path with defects or delay line or 30 mm straight path, respectively, while $\varphi_{(20\text{ mm})}$ represents the relative phase from 20 mm straight path waveguide as a reference. Figure 4 (f) shows that the topological waveguides with the same length have roughly similar time delays difference. It should be noted that the delay line sample is packaged on 20 mm straight waveguide length (same length), while 30 mm straight waveguide is just shown to compare the delay in time. Our results indeed demonstrate that the topologically protected propagation is robust to perturbations such as absence of pillars and sharp bends. These results can establish a path toward building a compact and low-cost 3D-printed topologically protected delay line capable of tens of picosecond delay.

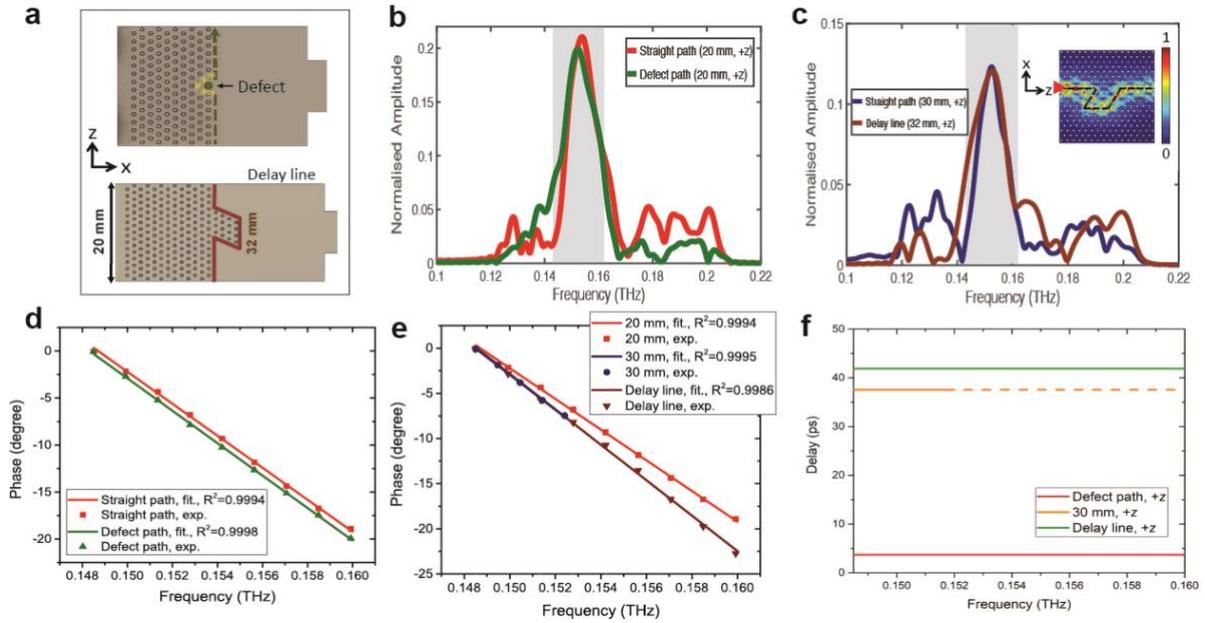

**Figure 4. THz robust topological experimental observation through defect path and delay line for forward (+z) propagation direction. a,** Schematics of the designed defect and delay line interfaces. Defect path: One pillar is removed from both metal plates in the middle of the channel path to create the defect. Delay line: interface with twisted path including four sharp bends (two 120° and two 60°). **b–c,** Experimentally measured amplitude spectra for THz radiations propagating along the straight path (20 mm) vs defect path (20 mm) and straight path (30 mm) vs delay line (32 mm) interface. **c,** The inset shows the field distribution in the delay line structure which demonstrates that light is strongly confined to 0.15 THz along the channel. The black line denotes the propagating direction of the sharply bent channel path. The red arrow shows that source is on the left side. **d,** The relative phases of straight and defect path in the bandgap region where the transmission is maximum (0.148 – 0.16 THz). **e,** The relative phases of straight paths and delay line. Note that 20 mm straight path phase is also inserted because the delay line is relative to 20 mm physical path length (**a**). 30 mm straight-path phase curve is only shown from 0.148 to 0.152 THz range due to high noise level. **f,** Measured time delays difference through defect, delay line and straight (30 mm) paths calculated using equation (1).

To conclude, we have designed, fabricated and characterized air-channel topologically protected THz waveguides. Both theoretical and experimental results confirmed topologically protected propagation over 0.143 – 0.162 THz (relative bandwidth of 12.5%), featuring relatively wide bandwidth, single-mode propagation and linearly guided modes. Moreover, our measurements demonstrate the robustness of the topological transport in the presence of defects and sharp bends. Compared to the all-silicon topological THz waveguides[46], our results show not only enhanced relative bandwidth (over 1.6 times increase compared to Ref. 46), but also reduced fabrication cost and complexity. Thanks to the rapid development in advanced 3D printing technology[47,51], we can envision more topologically protected photonic waveguide devices with fascinating functionalities in the THz band, having advantages of large bandwidth, low cost, robust transport, and excellent integration capacity inside the waveguide chip. Our work offers a promising topological integrated photonic platform towards high-speed, broadband THz signal transmission for next-generation short-range wireless communication.

**References**

1.  Tonouchi, M. Cutting-edge terahertz technology. *Nature Photonics* **1,** 97-105 (2007).

2.  Ma, J. *et al.* Security and eavesdropping in terahertz wireless links. *Nature* **563,** 89-93 (2018).

3.  Nagatsuma, T., Ducournau, G., Renaud, C. C. Advances in terahertz communications accelerated by photonics. *Nature Photonics* **10,** 371-379 (2016).

4.  Andrews, S. R. Microstructured terahertz waveguides. *Journal of Physics D: Applied Physics* **47,** 374004 (2014).

5.  Sengupta, K., Nagatsuma, T., Mittleman, D. M. Terahertz integrated electronic and hybrid electronic–photonic systems. *Nature Electronics* **1,** 622-635 (2018).

6.  Frankel, M. Y., Gupta, S., Valdmanis, J. A., Mourou, G. A. Terahertz attenuation and dispersion characteristics of coplanar transmission lines. *IEEE Trans. Microwave Theory Tech.* **39,** 910-916 (1991).

7.  Zhu, H., Xue, Q., Hui, J., Pang, S. W. Design, Fabrication, and Measurement of the Low-Loss SOI-Based Dielectric Microstrip Line and its Components. *IEEE Transactions on Terahertz Science and Technology* **6,** 696-705 (2016).

8.  Mendis, R., Mittleman, D. M. Comparison of the lowest-order transverse-electric (TE1) and transverse-magnetic (TEM) modes of the parallel-plate waveguide for terahertz pulse applications. *Optics Express* **17,** 14839-14850 (2009).

9.  Mendis, R., Grischkowsky, D. THz interconnect with low-loss and low-group velocity dispersion. *IEEE Microwave and Wireless Components Letters* **11,** 444-446 (2001).

10. Amarloo, H., Safavi-Naeini, S. Terahertz Line Defect Waveguide Based on Silicon-on-Glass Technology. *IEEE Transactions on Terahertz Science and Technology* **7,** 433-439 (2017).

11. Ranjkesh, N., Basha, M., Taeb, A., Safavi-Naeini, S. Silicon-on-Glass Dielectric Waveguide—Part II: For THz Applications. *IEEE Transactions on Terahertz Science and Technology* **5,** 280-287 (2015).

12. Bingham, A. L., Grischkowsky, D. R. Terahertz 2-D Photonic Crystal Waveguides. *IEEE Microwave and Wireless Components Letters* **18,** 428-430 (2008).

13. Li, H. *et al.* Broadband Single-Mode Hybrid Photonic Crystal Waveguides for Terahertz Integration on a Chip. *Advanced Materials Technologies* **5,** 2000117 (2020).

14. Tsuruda, K., Fujita, M., Nagatsuma, T. Extremely low-loss terahertz waveguide based on silicon photonic-crystal slab. *Optics Express* **23,** 31977-31990 (2015).

15. Atakaramians, S., Afshar V, S., Monro, T. M., Abbott, D. Terahertz dielectric waveguides. *Advances in Optics and Photonics* **5,** 169-215 (2013).

16. Li, H. *et al.* Flexible single-mode hollow-core terahertz fiber with metamaterial cladding. *Optica* **3,** 941-947 (2016).

17. Wang, K., Mittleman, D. M. Metal wires for terahertz wave guiding. *Nature* **432,** 376-379 (2004).

18. Yang, J. *et al.* 3D printed low-loss THz waveguide based on Kagome photonic crystal structure. *Optics Express* **24,** 22454-22460 (2016).

19. Joannopoulos, J. D., Johnson, S. G., Winn, J. N., Meade, R. D. *Photonic Crystals: Molding the Flow of Light*, 2nd edn, (Princeton University Press, Princeton, NJ, 2008).

20. Ma, T., Khanikaev, A. B., Mousavi, S. H., Shvets, G. Guiding Electromagnetic Waves around Sharp Corners: Topologically Protected Photonic Transport in Metawaveguides. *Phys. Rev. Lett.* **114,** 127401 (2015).

21. Bernevig, B. A., Zhang, S.-C. Quantum Spin Hall Effect. *Phys. Rev. Lett.* **96,** 106802 (2006).

22. Hasan, M. Z., Kane, C. L. Colloquium: Topological insulators. *Rev. Mod. Phys.* **82,** 3045-3067 (2010).

23. Khanikaev, A. B., Shvets, G. Two-dimensional topological photonics. *Nature Photonics* **11,** 763-773 (2017).

24. Haldane, F. D. M., Raghu, S. Possible Realization of Directional Optical Waveguides in Photonic Crystals with Broken Time-Reversal Symmetry. *Phys. Rev. Lett.* **100,** 013904 (2008).

25. Wang, Z., Chong, Y., Joannopoulos, J. D., Soljačić, M. Observation of unidirectional backscattering-immune topological electromagnetic states. *Nature* **461,** 772-775 (2009).

26. Hafezi, M., Demler, E. A., Lukin, M. D., Taylor, J. M. Robust optical delay lines with topological protection. *Nature Physics* **7,** 907-912 (2011).

27. Kraus, Y. E., Lahini, Y., Ringel, Z., Verbin, M., Zilberberg, O. Topological States and Adiabatic Pumping in Quasicrystals. *Phys. Rev. Lett.* **109,** 106402 (2012).

28. Fang, K., Yu, Z., Fan, S. Realizing effective magnetic field for photons by controlling the phase of dynamic modulation. *Nature Photonics* **6,** 782-787 (2012).

29. Khanikaev, A. B., Hossein Mousavi, S., Tse, W.-K., Kargarian, M., MacDonald, A. H., Shvets, G. Photonic topological insulators. *Nature Materials* **12,** 233-239 (2013).

30. Rechtsman, M. C. *et al.* Photonic Floquet topological insulators. *Nature* **496,** 196-200 (2013).

31. Bahari, B., Ndao, A., Vallini, F., El Amili, A., Fainman, Y., Kanté, B. Nonreciprocal lasing in topological cavities of arbitrary geometries. *Science* **358,** 636-640 (2017).

32. Harari, G. *et al.* Topological insulator laser: Theory. *Science* **359,** eaar4003 (2018).

33. Guglielmon, J., Rechtsman, M. C. Broadband Topological Slow Light through Higher Momentum-Space Winding. *Phys. Rev. Lett.* **122,** 153904 (2019).

34. Barik, S. *et al.* A topological quantum optics interface. *Science* **359,** 666-668 (2018).

35. Blanco-Redondo, A., Bell, B., Oren, D., Eggleton, B. J., Segev, M. Topological protection of biphoton states. *Science* **362,** 568-571 (2018).

36. Mittal, S., Goldschmidt, E. A., Hafezi, M. A topological source of quantum light. *Nature* **561,** 502-506 (2018).

37. Wang, M. *et al.* Topologically protected entangled photonic states. *Nanophotonics* **8,** 1327 (2019).

38. Blanco-Redondo, A. Topological Nanophotonics: Toward Robust Quantum Circuits. *Proc. IEEE* **108,** 837-849 (2020).

39. Lai, K., Ma, T., Bo, X., Anlage, S., Shvets, G. Experimental Realization of a Reflections-Free Compact Delay Line Based on a Photonic Topological Insulator. *Scientific Reports* **6,** 28453 (2016).

40. Yves, S., Fleury, R., Berthelot, T., Fink, M., Lemoult, F., Lerosey, G. Crystalline metamaterials for topological properties at subwavelength scales. *Nature Communications* **8,** 16023 (2017).

41. Chen, Q. *et al.* Robust waveguiding in substrate-integrated topological photonic crystals. *Appl. Phys. Lett.* **116,** 231106 (2020).

42. Bandres, M. A. *et al.* Topological insulator laser: Experiments. *Science* **359,** eaar4005 (2018).

43. Hafezi, M., Mittal, S., Fan, J., Migdall, A., Taylor, J. M. Imaging topological edge states in silicon photonics. *Nature Photonics* **7,** 1001-1005 (2013).

44. He, X.-T. *et al.* A silicon-on-insulator slab for topological valley transport. *Nature Communications* **10,** 872 (2019).

45. Shalaev, M. I., Walasik, W., Tsukernik, A., Xu, Y., Litchinitser, N. M. Robust topologically protected transport in photonic crystals at telecommunication wavelengths. *Nature Nanotechnology* **14,** 31-34 (2019).

46. Yang, Y. *et al.* Terahertz topological photonics for on-chip communication. *Nature Photonics* **14,** 446-451 (2020).

47. del Barrio, J., Sánchez-Somolinos, C. Light to Shape the Future: From Photolithography to 4D Printing. *Advanced Optical Materials* **7,** 1900598 (2019).

48. Gentili, E., Tabaglio, L., Aggogeri, F. Review on Micromachining Techniques. AMST'05 Advanced Manufacturing Systems and Technology: Springer 2005. p. 387-396.

49. Ligon, S. C., Liska, R., Stampfl, J., Gurr, M., Mülhaupt, R. Polymers for 3D Printing and Customized Additive Manufacturing. *Chem. Rev.* **117,** 10212-10290 (2017).

50. Wu, Z., Ng, W.-R., Gehm, M. E., Xin, H. Terahertz electromagnetic crystal waveguide fabricated by polymer jetting rapid prototyping. *Optics Express* **19,** 3962-3972 (2011).

51. Sun, J., Hu, F. Three-dimensional printing technologies for terahertz applications: A review. *International Journal of RF and Microwave Computer-Aided Engineering* **30,** e21983 (2020).

52. Rahiminejad, S., Köhler, E., Enoksson, P. Direct 3D printed shadow mask on Silicon. Journal of Physics: Conference Series: IOP Publishing; 2016. p. 012021.

53. Yudasari, N., Anthony, J., Leonhardt, R. Terahertz pulse propagation in 3D-printed waveguide with metal wires component. *Optics Express* **22,** 26042-26054 (2014).

54. Zhang, B., Guo, Y., Zirath, H., Zhang, Y. P. Investigation on 3-D-Printing Technologies for Millimeter-Wave and Terahertz Applications. *Proc. IEEE* **105,** 723-736 (2017).

55. Xin, H., Liang, M. 3-D-Printed Microwave and THz Devices Using Polymer Jetting Techniques. *Proc. IEEE* **105,** 737-755 (2017).

56. Combrié, S. *et al.* Detailed analysis by Fabry-Perot method of slab photonic crystal line-defect waveguides and cavities in aluminium-free material system. *Optics Express* **14,** 7353-7361 (2006).

**Acknowledgments**

Sputtering of the samples were performed in part at the NSW Node of the Australian National Fabrication Facility in UNSW and University of Sydney Research and Prototype foundry. The 3D samples were printed at School of Mechanical and Manufacturing Engineering laboratory, UNSW. H.L. acknowledges research funding support from the National Natural Science

Foundation of China (grant number 62075007), and the Beijing Natural Science Foundation, China (grant number 4192048). S.A. acknowledges the support of UNSW Scientia program.

**Author Contributions**

All the authors contributed to the data analysis, interpretation of the results and writing of manuscript. H.L., A.B.-R. and S.A. performed simulations. N.N.M.D., H.L., A.B.-R. and S.A. worked on the design and fabrication of waveguides. M.T.A.K., N.N.M.D., H.L. and S.A. performed the experiment. A.B.-R. and S.A. conceived the idea. S.A. mapped out the steps of the project and supervised the project.

**Additional Information**

The authors declare no conflicts of interest. Correspondence and requests for materials should be addressed to Muhammad Talal Ali Khan (m.t.khan@unsw.edu.au), Haisu Li (lihaisu@bjtu.edu.cn) and Shaghik Atakaramians (s.atakaramians@unsw.edu.au).